\documentclass{sigchi}

\toappear{\scriptsize Permission to make digital or hard copies of all or part of this work for personal or classroom use is granted without fee provided that copies are not made or distributed for profit or commercial advantage and that copies bear this notice and the full citation on the first page. Copyrights for components of this work owned by others than ACM must be honored. Abstracting with credit is permitted. To copy otherwise, or republish, to post on servers or to redistribute to lists, requires prior specific permission and/or a fee. Request permissions from permissions@acm.org. \\
{\emph{IUI 2018, March 7--11, 2018, Tokyo, Japan.} } \\
Copyright \copyright~2018 ACM ISBN 978-1-4503-4945-1/18/03\ ...\$15.00. \\
http://dx.doi.org/10.1145/XXX.XXXXX}
\usepackage{balance}       % to better equalize the last page
\usepackage{graphics}      % for EPS, load graphicx instead 
\usepackage[T1]{fontenc}   % for umlauts and other diaeresis
\usepackage{txfonts}
\usepackage{mathptmx}
\usepackage[pdflang={en-US},pdftex]{hyperref}
\usepackage{color}
\usepackage{booktabs}
\usepackage{textcomp}
\usepackage[export]{adjustbox}

\usepackage{microtype}        % Improved Tracking and Kerning
\usepackage{ccicons}          % Cite your images correctly!
\usepackage{todonotes}

\def\plaintitle{Coupling Story to Visualization: Using Textual Analysis as a Bridge Between Data and Interpretation}

\def\emptyauthor{}
\def\plainkeywords{Authors' choice; of terms; separated; by
  semicolons; include commas, within terms only; required.}

\makeatletter
\def\url@leostyle{%
  \@ifundefined{selectfont}{
    \def\UrlFont{\sf}
  }{
    \def\UrlFont{\small\bf\ttfamily}
  }}
\makeatother
\def\pprw{8.5in}
\def\pprh{11in}

\definecolor{linkColor}{RGB}{6,125,233}
\hypersetup{%
  pdftitle={\plaintitle},
  pdfauthor={\emptyauthor},
  pdfkeywords={\plainkeywords},
  pdfdisplaydoctitle=true, % For Accessibility
  bookmarksnumbered,
  pdfstartview={FitH},
  colorlinks,
  citecolor=black,
  filecolor=black,
  linkcolor=black,
  urlcolor=linkColor,
  breaklinks=true,
  hypertexnames=false
}

\begin{document}

\title{\plaintitle}

\numberofauthors{1}
% \author{%
%   \alignauthor{Ronald Metoyer\\
%     \affaddr{University of Notre Dame}\\
%     \email{rmetoyer@nd.edu}}\\
%   \alignauthor{Walter Scheirer\\
%     \affaddr{University of Notre Dame}\\
%     \email{walter.scheirer@nd.edu}}\\
%   \alignauthor{Qiyu Zhi\\
%     \affaddr{University of Notre Dame}\\
%     \email{qzhi@nd.edu}}\\
%   \alignauthor{Bart Janczuk\\
%     \affaddr{University of Notre Dame}\\
%     \email{bjanczuk@nd.edu}}\\
% }

\author{%
  \alignauthor{Ronald Metoyer, Qiyu Zhi, Bart Janczuk, Walter Scheirer\\
    \affaddr{University of Notre Dame}\\
    \affaddr{Notre Dame, USA}\\
    \email{\{rmetoyer, qzhi, bjanczuk, walter.scheirer\}@nd.edu}}\\
}

\maketitle

\begin{abstract}

 Online writers and journalism media are increasingly combining visualization (and other multimedia content) with narrative text to create narrative visualizations.  Often, however, the two elements are presented independently of one another.  We propose an approach to automatically integrate text and visualization elements.  We begin with a writer's narrative that presumably can be supported with visual data evidence.  We leverage natural language processing, quantitative narrative analysis, and information visualization to (1) automatically extract narrative components (who, what, when, where) from data-rich stories, and (2) integrate the supporting data evidence with the text to develop a narrative visualization.   We also employ bidirectional interaction from text to visualization and visualization to text to support reader exploration in both directions.  We demonstrate the approach with a case study in the data-rich field of sports journalism.
  
\end{abstract}

\category{}{Human-centered computing}{Information visualization} \category{}{Computing methodologies}{Information extraction}{}
  

\section{Introduction}

Integrating narrative text with visual evidence is a fundamental aspect of various investigative reporting fields such as sports journalism and intelligence analysis.  Sports journalists generate stories about sporting events that typically generate tremendous amounts of data.  These stories summarize the event and build support for their particular narrative (e.g. why a particular team won or lost) using the data as evidence.    Intelligence analysts carry out similar tasks, collecting data as evidence in support of the narrative that they construct to inform decision makers~\cite{arcos2015intelligence}. Such a process would also be familiar to digital humanities scholars who routinely distill meta-narratives by automatic means from large corpora consisting of literary texts~\cite{moretti2013distant,jockers2013macroanalysis}. Support for linking visualizations of this information to the text is an underexplored area that could benefit each of these users.

\begin{figure}[t!]
\centering
\includegraphics[width=.48\columnwidth, frame]{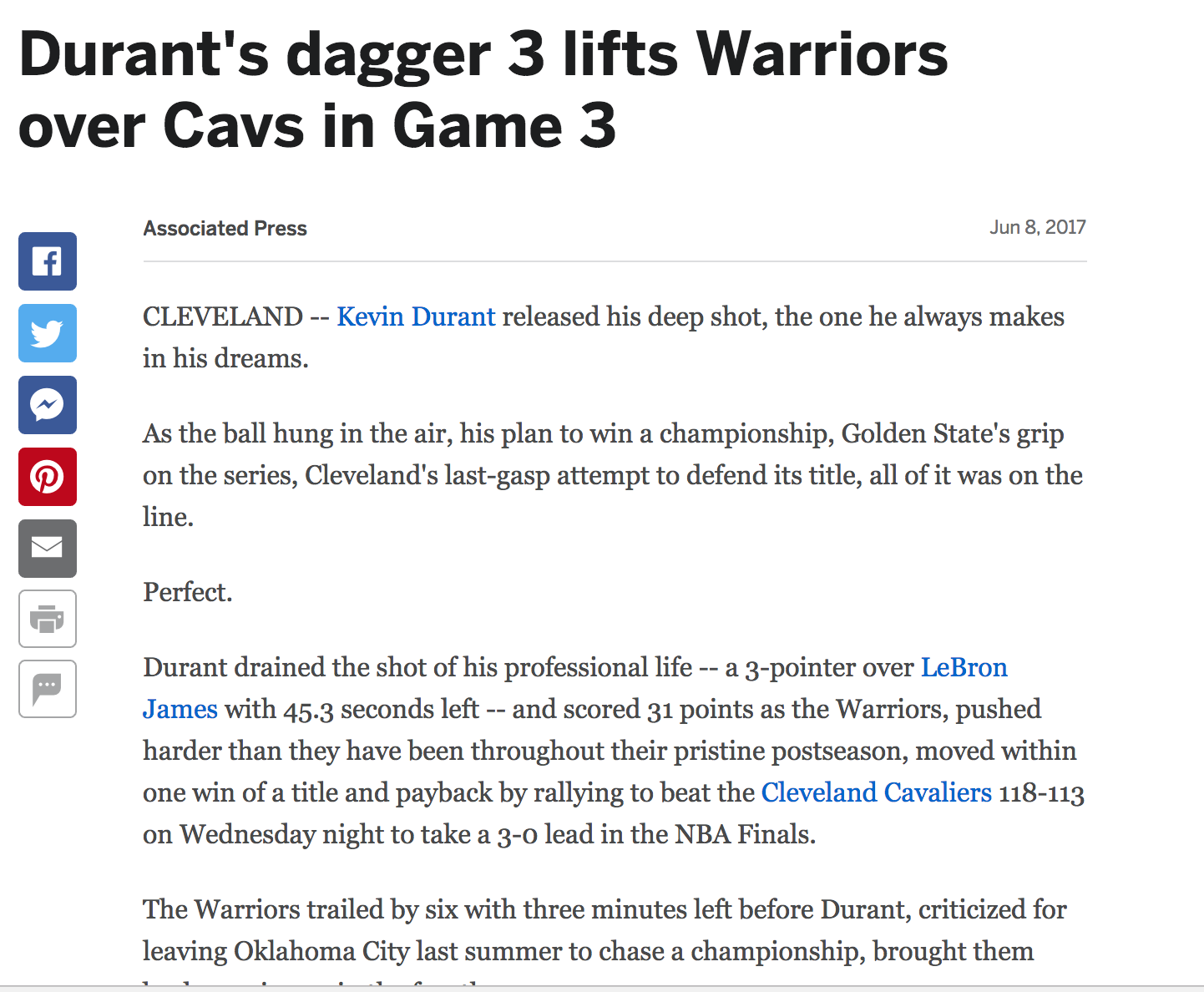}
\includegraphics[width=.48\columnwidth, frame]{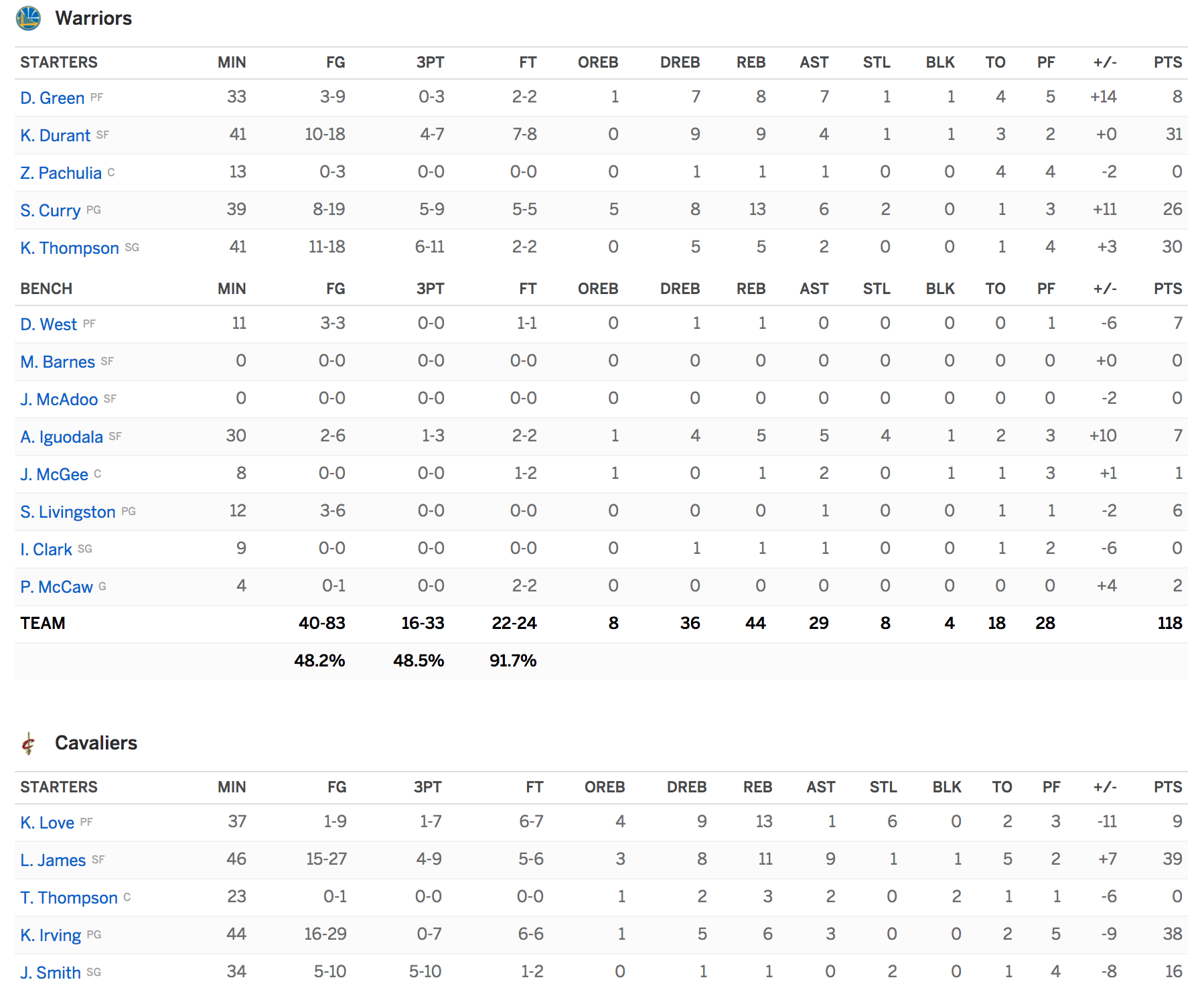}
\includegraphics[width=.48\columnwidth, frame]{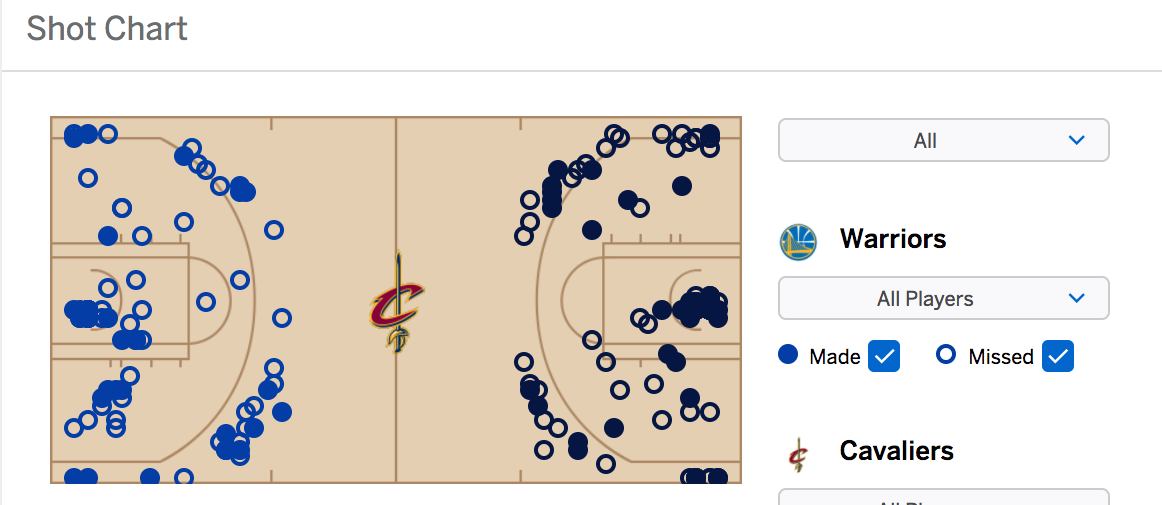}
\includegraphics[width=.48\columnwidth, frame]{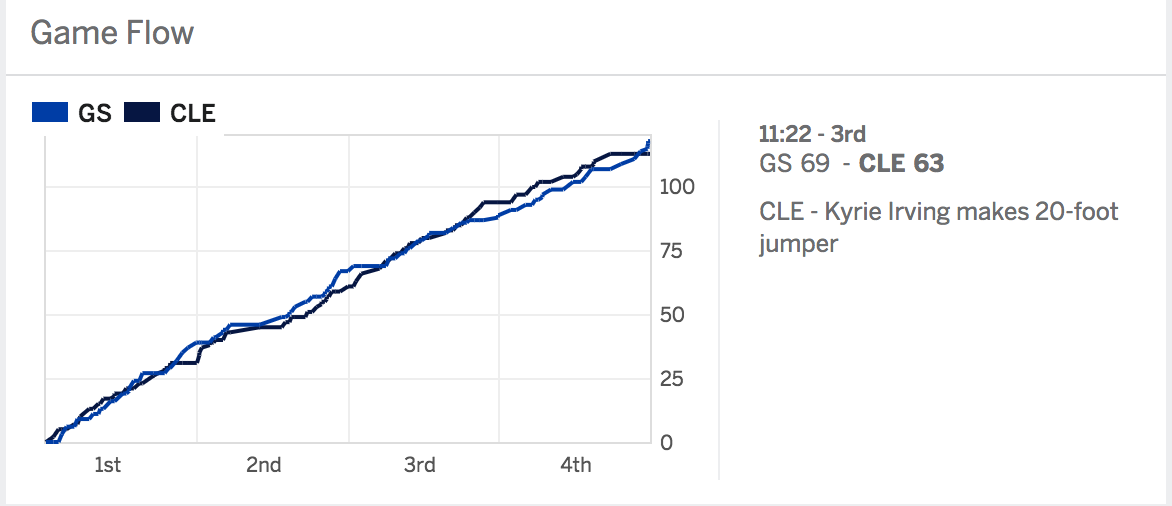}
\caption{Sites such as www.espn.com provide the reader with several forms of content, but they are often presented independent of one another.  Here we see typical narrative text on the upper left, box score in the upper right, and visualizations (shot chart and game flow time series) along the bottom.  Each of these elements is shown on a separate web page on the ESPN site. All content from www.espn.com}
\label{fig:espn}
\end{figure}

When visual evidence is provided, it is often presented completely independently of the narrative text. 
 ESPN\footnote{espn.go.com}, for example, routinely presents journalistic articles on web pages that are separate from relevant videos, interactive visualizations, and box score tables that all present information about the same game event (See Figure \ref{fig:espn}).  This approach may limit the reader's ability to interpret the writer's narrative in context.
 For example, a reader might be interested in more detail surrounding a specific element described in the narrative text because she may not agree with the statement.  Alternatively, the reader might also be interested in finding the writer's particular take on a specific player of interest or time in the game that she came across while exploring the data visualization.  This bi-directional support for interpretation is currently lacking. We leverage the existing skill set of the writer and combine it with computational methods to automatically link the writer's narrative text to visualization elements in order to provide that context in both directions.

\section{Related Work}
Narrative visualizations are data visualizations with embedded ``stories'' presenting particular perspectives using various embedding mechanisms\cite{segel2010narrative}.  Segel et al. \cite{segel2010narrative} discuss the utility of textual ``messaging''and how author-driven approaches typically include heavy messaging while reader-driven approaches involve more interaction.  Kosara and Mackinlay \cite{kosara2013storytelling} and Lee et al. \cite{lee2015more}, on the other hand, adopt a view of storytelling in visualization as primarily consisting of communication separate from exploratory analysis. This view matches the way journalists use various data sources including videos and notes to create compelling stories \cite{kosara2013storytelling}.

We attempt to provide equal flexibility to the reader and author. Our proposed method for coupling textual stories to data visualizations can not only assist writers in constructing their message and framing the data \cite{hullman2011visualization, kosara2013storytelling} but also create exploratory interaction mechanisms for readers. The linking of subjective narrative with objective data evidence can serve to reinforce author communication and give readers the ability to explore their own interpretations of the story and the event. This approach can be viewed as providing a ``close reading'' of the narrative, using the visualization as a means for providing the semantic context \cite{janicke2015close} as opposed to a distant reading where the actual text is replaced by abstract visual views \cite{moretti2013distant}.

We are not the first to investigate the coupling of text to visualization.
Chul Kwon et al. \cite{kwon2014visjockey} directly, but manually, tie specific text elements to visualization views to support interpretation of the author's content.  Others analyze text to automatically generate elements of narrative visualization such as annotations \cite{hullman2013contextifier, bryan2017temporal}. Likewise, Dou et al. \cite{dou2012leadline} also analyze text, but from large scale text corpora, to identify key ``event'' elements of who, what, when and where and use that information to construct a visual summary of the corpora and support exploration of it.  We build on these approaches, automatically linking text to visualization elements, but focus on a single data-rich text document.  In addition, we explore the use of links from the visualization back to the text.

\section{Sports Stories: Basketball}
To ground the discussion, we focus on the data-rich domain of sports, specifically basketball, for which we have a significant set of narratives and data published regularly online. Sports events produce tremendous amounts of data and are the subject of sports journalists and general public fans.  Basketball is no exception.  Data regarding player position and typical basketball  ``statistics'' are collected for every game in the National Basketball Association (NBA)\footnote{http://stats.nba.com/}.  These data typically include field goals (2pt and 3pt), assists, rebounds, steals, blocked shots, turnovers, fouls, and minutes played.   The NBA has recently begun to track the individual positions of players throughout the entire game, leading to spatial data as well as various derived data such as the total number of times a player touches the ball  (i.e., touches). The data is rich in space, time, and discrete nominal and quantitative attributes.  

\section{Overview}
\begin{figure}%[l]{0.9\textwidth}
%\centering
\begin{center}
\includegraphics[width=0.50\textwidth]{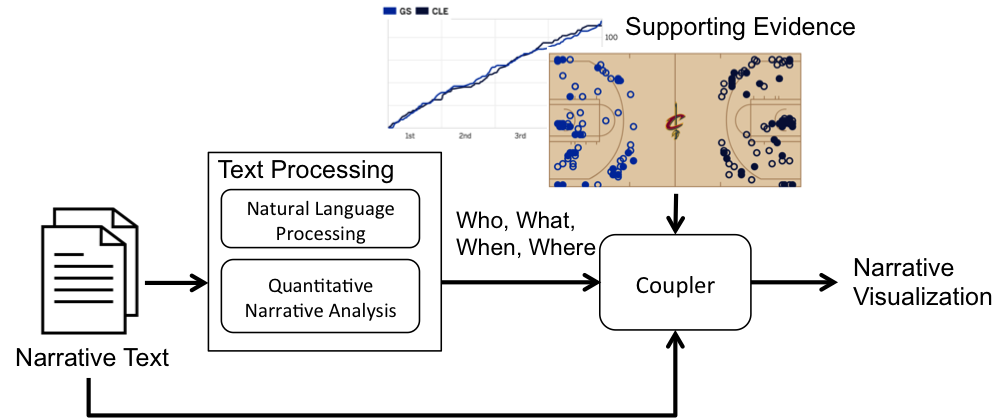}
\end{center}
\caption{Software architecture for deep coupling of narrative text and visualization components.}
\label{fig:overview}
\end{figure}

Our approach is based on a detailed analysis of narrative text to identify key elements of the story that can directly index into the visualizations (and vice versa) to create a \textit{deep coupling}. The system architecture is shown in Figure \ref{fig:overview}.  Raw text is provided as input to the text processing module.  This module analyzes the text to extract the four key narrative elements - who, what, when, and where (4 Ws).  The coupler then indexes into the multimedia and vice versa with these Ws. We focus solely on data visualization multimedia elements (charts, tables, etc.) as the supporting evidence, however, evidence such as video, images, and audio could be treated similarly.

\section{Identifying Story Elements}

A story is defined by characters, plot, setting, theme, and goal which are also sometimes referred to as the five Ws: who, what, when, where and why.  These are the elements considered necessary to tell and/or uncover, in the case of intelligence investigation,  a complete story \cite{arcos2015intelligence}.  

In sports, there are clear mappings for these essential elements.  In basketball, for example, there are players, coaches, referees, and teams (who), locations/spaces on the basketball court such as the 3-point line or inside the key (where), distinct times or time periods such as the 4th quarter, or 3-minutes left (when), and particular game ``stats'' such as shots, fouls, timeouts, touches, etc. (what).  Our first step is to extract these elements.

We first segment the text into sentences and for each sentence, we seek to determine if it contains a reference to who, what, when, or where. Consider the following sentence from an Associated Press story from Game 3 of the 2017 NBA Finals between the Cleveland Cavaliers and the Golden State Warriors \footnote{http://www.espn.com/nba/recap?gameId=400954512}:\textit{
The Warriors trailed by six with three minutes left before Durant, criticized for leaving Oklahoma City last summer to chase a championship, brought them back, scoring 14 in the fourth.}  From the second clause in this sentence, we would expect to extract a Who (Durant), a What (14 points), a When (the fourth), and no Where elements.

\noindent \textbf{Who:} 
Extracting the story characters is straightforward, especially for a particular domain.  In the NBA case, we can obtain a complete list of all teams, players, coaches, and referees in the league. A more general solution, however,  could utilize a named-entity identifier such as the Stanford Core NLP Named Entity Recognizer \footnote{https://nlp.stanford.edu/software/CRF-NER.shtml}. 

\noindent \textbf{What:} The ``What'' of a basketball story, or any sports story in general, refers to the game statistics of interest. While it would seem straightforward to look for mentions of numbers and/or specific stat names since numbers and names are generally used to describe these stats (e.g. 25 points, 10 rebounds,  five 3-pointers), this is not always the case.  Consider the many ways to refer to the points scored stat: Durant scored 25 points, Durant added 15, or even, Durant outscored the entire Cavalier team in the second quarter.  

To identify stats, we take inspiration from the digital humanities  \cite{franzosi2010quantitative} and build a ``story grammar'' to parse the narrative.  The stats story grammar is a set of manually designed production rules for creating and/or matching variations of mentions of stats. While the grammar-based approach is appropriate for identifying most mentions of stats, it is only as effective as the grammar specification and a new variation of a stat mention will not be recognized.  We therefore use the grammar to generate training data for an SVM classifier for stats.

We run the sentences of 13 stories/blog posts through the stats story grammar to label the sentences as STAT or NO STAT.  We then use these positive (STAT) sentence-classification pairs as training data for the SVM classifier.  Specifically, for each sentence that contains a statistic, we create a 10-word window around the statistic to use as a feature vector.  All such windows are provided as training examples in order to learn the general textual context that surrounds mentions of stats.  After training on approximately 600 such windows, the classifier achieves 98\% accuracy on a test data set of approximately 150 sentences and it finds additional statistic mentions that were not classified correctly by the grammar.  

\noindent \textbf{When and Where:}  We currently extract both times and locations using only a grammar with acceptable results because there are a limited number of ways to specify times and places in basketball.  A classifier similar to that developed for stats could also be generated for both times and places to boost performance if necessary.

\subsection{Narrative Text to Visualization Coupling}
The text analysis stage results in a list of Ws for each sentence in the narrative, stored as a .json object with W-value pairs.
This information is then provided to the ``coupling'' component.  The coupler associates each sentence with the appropriate visualization(s) (i.e., those that support the specific Ws in the sentence) and initializes the visualization to show the data for the particular person (who), stat (what), time (when), and location (where) mentioned in the sentence. 

In Figure \ref{fig:annotated}, the user has hovered over the sentence highlighted on the left side. The mentioned player is highlighted in the box score on the right (Kevin Durant) and his player profile is shown at the bottom left.  The specific mentioned time in the game is marked in the time series visualization (3 minutes left), the specific quarter that is mentioned is selected in the time series visualization (fourth quarter), and the shot chart in the upper left of the visualization shows the shots attempted (both made and missed) by Kevin Durant in the 4th quarter.   This visualization initialization provides the context for the writer's interpretation of what was important, and allows the reader to ask their own questions, such as: How many shots did Durant have to take in the 4th quarter to score 14 points?

\subsection{Interaction Design}
No guidelines or principles currently exist for the design of tightly coupled reader interaction with narrative text and visualization concurrently. We chose a multi-view or ``partitioned poster'' layout \cite{segel2010narrative} to allow for all visualizations to be accessible in a single view along with the narrative text.  Interaction is guided by basic multi-view coordination principles \cite{north2000snap}.  We assume each data table consists of items described by attributes - both of which can correspond directly to one of the four Ws of a story. Thus a visualization can be directly parameterized by one or more of the 4Ws from a sentence.   

\begin{figure*}[h!]
\centering
\includegraphics[width=0.82\linewidth]{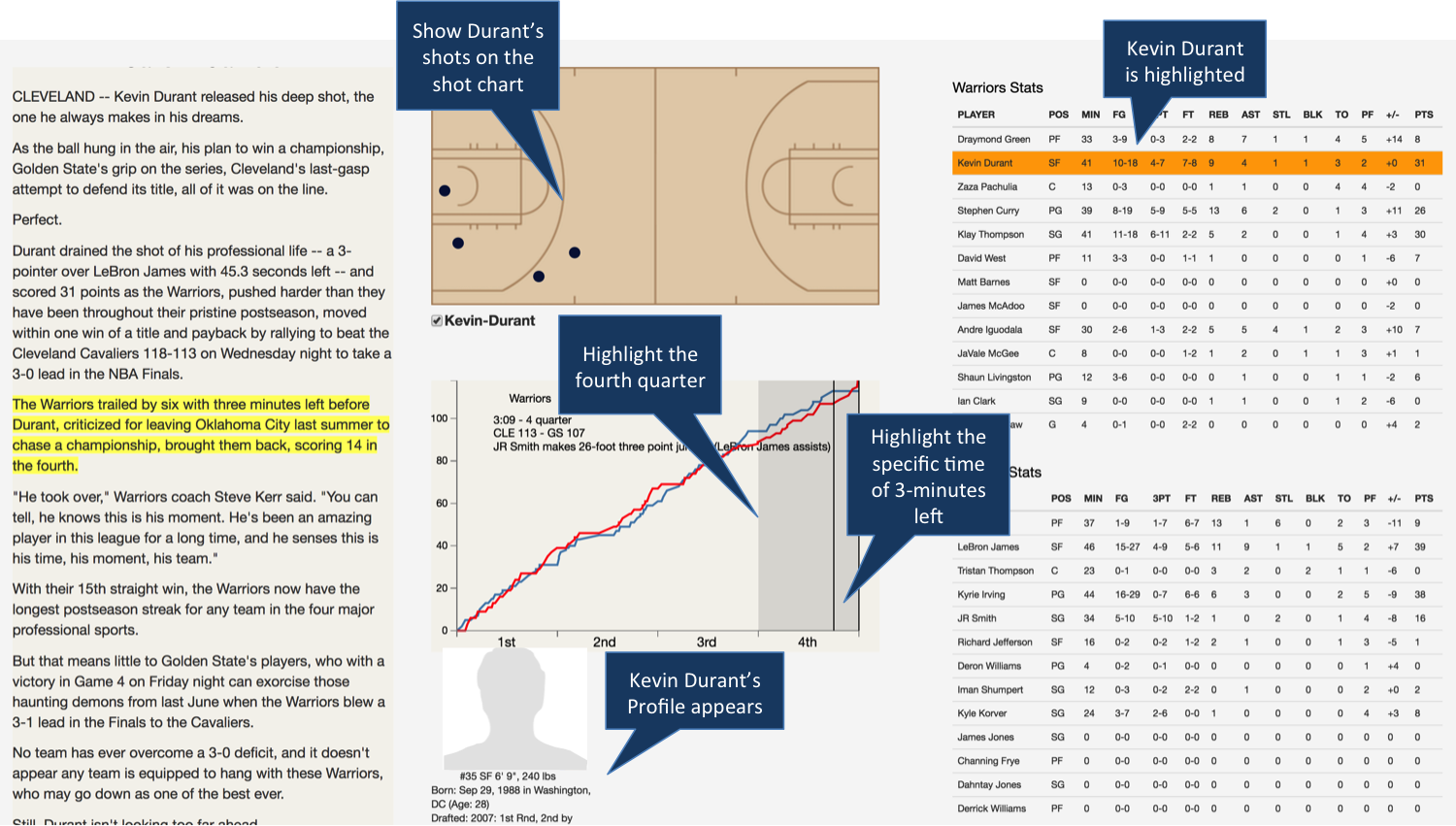}
\caption{Coupling from text to visualization.  The user has selected a passage that reads:  The Warriors trailed by six with three minutes left before Durant, criticized for leaving Oklahoma City last summer to chase a championship, brought them back, scoring 14 in the fourth. Story excerpt from the Associated Press. }
\label{fig:annotated}
\end{figure*}

 We coordinate text views with visualization views and vice versa.  User interaction  (e.g., mouse over, selection) with specific sentences results in the corresponding Ws being brushed in the linked visualizations.  Likewise, user interaction with the visualization components results in the corresponding Ws being highlighted in the narrative text.    For example, in Figure \ref{fig:annotated}, users can select a row (i.e. a player) in the box score tables and that player's data is then highlighted (brushing and linking) in all other views (e.g. shots in the short chart and mentions in the time series).  That player's profile is also displayed in detail at the bottom left (details on demand).   All visualizations support tooltip details where appropriate.

\subsubsection{Navigation Between Text and Visualization}
The reader must be able to navigate from the narrative text to the visual representations and vice versa.   As the reader browses the text, sentences are highlighted to show the current sentence of interest and the visualizations update to reflect the corresponding Ws from the highlighted text. The user can click on a sentence to indicate that he is interested in exploring the data context around that aspect of the narrative, thus, we freeze the state of the visualizations.
%(i.e., the parameterization by Ws).  
The user can then interact with the visualizations, using mouse-overs to get further details. Any direct interaction, such as a click on an element or brush to select a time series range, results in updating all visualizations and the highlighted narrative text given the new selected Ws. The user can now continue to explore the visualizations or can navigate back to the narrative text at any time to examine the writer's commentary with regards to the selected Ws.

\section{Case Study: Sports Narratives}

To gain an initial understanding of the potential opportunities and pitfalls that this approach presents, we conducted an informal case study with a domain expert user with 4.5 years of experience as a media producer for Bleacher Report\footnote{www.bleacherreport.com} and multiple collegiate sports media outlets.  We gave the user a brief introduction to the interface and asked him to explore a game recap of Game 3 of the 2017 NBA finals and provide feedback using a ``Think Aloud'' protocol.  

Our user immediately grasped the navigation concept and was easily able to move from reading/scanning the narrative text to the visualization and back to the narrative text.   His general impression was that there was great value in being able to more deeply explore the specific narrative presented by the author:   \textit{I love the idea of being able to experience the article, but really dial-in on a specific part of it myself.}   He was also intrigued by the potential to more deeply tie the text phrasing to the visuals. \textit{If the writer emphasized the number of points Curry had from the paint, I'd be interested in seeing that visually, and also exploring for myself where else he may have scored from. I'd like to select areas like the paint, midrange, and 3-pointers and beyond to see the shots he took in those areas.}

Our user also articulated several areas for improvement.  For example, he found the box score filtering by selecting a player row to be extremely useful, but wanted even more control.  \textit{I'd like to select both a player row and a stat column so I could more specifically look for a specific stat for that player in the other charts like the time series view.} He also recommended a more subtle tie from the visualization to the narrative text might be more effective.  Currently, data items selected through direct interaction with the visualization causes the sentences that mentioned those items to be highlighted in the narrative.  This was sometimes overwhelming because some sentences may have directly mentioned specific players and their stats while others may have simply provided commentary or mentioned the player in a secondary role.  A deeper analysis of the text to understand the saliency of the sentence with regards to the selected data should be explored. Finally, our user desired a more granular link to the text.  \textit{It would be useful if the text highlighted the types of elements - player, stats, times, etc., with different colors }.    Future versions could employ color encodings to distinguish the 4Ws in the text.

\section{Conclusion and Future Work}
We have presented a novel approach to automatically couple narrative text to visualizations for data-rich stories and demonstrated the approach in the basketball story domain.  
%The approach, however, is generally applicable to any narrative written about data-rich events. 
This serves only as an introductory step for using text analysis to tightly integrate visualization with narrative text. There are several exciting avenues of research that remain to be explored. 

First, Where is the 5th W?  We have not attempted to analyze the text to understand the ``Why'' of the narrative. A more nuanced narrative analysis is necessary to take semantics and topics into account when coupling to the visualization.

Our approach is carried out completely offline.  One can imagine, however, a real-time version that operated online as the journalist writes the story - suggesting a set of initialized candidate visualizations to support the most recently completed sentence. Our case study user was intrigued by the possibility that the tool could influence how journalists write. \textit{I wonder if a  writer might use specific phrasing as a tool for leading the reader to particular elements of the game?}  It would be interesting to study how real-time interactive visualization coupling might influence the writing style of journalists.

This approach should transfer well to any sport where stories typically consist of discussions of players, teams, the field (or pitch, court, etc.), times, and game stats.  We suspect it can also be extended to data-rich news stories as well, however, these will require more sophisticated algorithms for extracting the Ws given the more general language and context.  Finally, a thorough evaluation of the approach is needed.  We are in the process of conducting a crowdsourced study to examine the effects of this type of coupling on the reading experience.
% BALANCE COLUMNS
\balance{}

% REFERENCES FORMAT
% References must be the same font size as other body text.
\bibliographystyle{SIGCHI-Reference-Format}
\bibliography{sample}


\begin{thebibliography}{00}

%%% the Web address.

\ifx \showCODEN    \undefined \def \showCODEN     #1{\unskip}     \fi
\ifx \showDOI      \undefined \def \showDOI       #1{{\tt DOI:}\penalty0{#1}\ }
  \fi
\ifx \showISBNx    \undefined \def \showISBNx     #1{\unskip}     \fi
\ifx \showISBNxiii \undefined \def \showISBNxiii  #1{\unskip}     \fi
\ifx \showISSN     \undefined \def \showISSN      #1{\unskip}     \fi
\ifx \showLCCN     \undefined \def \showLCCN      #1{\unskip}     \fi
\ifx \shownote     \undefined \def \shownote      #1{#1}          \fi
\ifx \showarticletitle \undefined \def \showarticletitle #1{#1}   \fi
\ifx \showURL      \undefined \def \showURL       #1{#1}          \fi

\bibitem{arcos2015intelligence}
{Ruben Arcos} {and} {Randolph Pherson}. 2015.
\newblock {\em {Intelligence Communication in the Digital Era: Transforming
  Security, Defence and Business}}.
\newblock Springer.
\newblock


\bibitem{bryan2017temporal}
{Chris Bryan}, {Kwan-Liu Ma}, {and} {Jonathan Woodring}. 2017.
\newblock \showarticletitle{Temporal summary images: An approach to narrative
  visualization via interactive annotation generation and placement}.
\newblock {\em IEEE transactions on visualization and computer graphics\/}
  {23}, 1 (2017), 511--520.
\newblock


\bibitem{dou2012leadline}
{Wenwen Dou}, {Xiaoyu Wang}, {Drew Skau}, {William Ribarsky}, {and} {Michelle~X
  Zhou}. 2012.
\newblock \showarticletitle{{Leadline: Interactive visual analysis of text data
  through event identification and exploration}}. In {\em Visual Analytics
  Science and Technology (VAST), 2012 IEEE Conference on}. IEEE, 93--102.
\newblock


\bibitem{franzosi2010quantitative}
{Roberto Franzosi}. 2010.
\newblock {\em {Quantitative narrative analysis}}.
\newblock Number 162. Sage.
\newblock


\bibitem{hullman2011visualization}
{Jessica Hullman} {and} {Nick Diakopoulos}. 2011.
\newblock \showarticletitle{{Visualization rhetoric: Framing effects in
  narrative visualization}}.
\newblock {\em IEEE transactions on visualization and computer graphics\/}
  {17}, 12 (2011), 2231--2240.
\newblock


\bibitem{hullman2013contextifier}
{Jessica Hullman}, {Nicholas Diakopoulos}, {and} {Eytan Adar}. 2013.
\newblock \showarticletitle{{Contextifier: automatic generation of annotated
  stock visualizations}}. In {\em Proceedings of the SIGCHI Conference on Human
  Factors in Computing Systems}. ACM, 2707--2716.
\newblock


\bibitem{janicke2015close}
{Stefan J{\"a}nicke}, {Greta Franzini}, {Muhammad~Faisal Cheema}, {and} {Gerik
  Scheuermann}. 2015.
\newblock \showarticletitle{{On close and distant reading in digital
  humanities: A survey and future challenges}}.
\newblock {\em Proc. EuroVis, Cagliari, Italy\/} (2015).
\newblock


\bibitem{jockers2013macroanalysis}
{Matthew~L. Jockers}. 2013.
\newblock {\em Macroanalysis: Digital Methods and Literary history}.
\newblock University of Illinois Press.
\newblock


\bibitem{kosara2013storytelling}
{Robert Kosara} {and} {Jock Mackinlay}. 2013.
\newblock \showarticletitle{{Storytelling: The next step for visualization}}.
\newblock {\em Computer\/} {46}, 5 (2013), 44--50.
\newblock


\bibitem{kwon2014visjockey}
{Bum~Chul Kwon}, {Florian Stoffel}, {Dominik J{\"a}ckle}, {Bongshin Lee}, {and}
  {Daniel Keim}. 2014.
\newblock \showarticletitle{VisJockey: Enriching Data Stories through
  Orchestrated Interactive Visualization}. In {\em Poster Compendium of the
  Computation+ Journalism Symposium}, Vol.~3.
\newblock


\bibitem{lee2015more}
{Bongshin Lee}, {Nathalie~Henry Riche}, {Petra Isenberg}, {and} {Sheelagh
  Carpendale}. 2015.
\newblock \showarticletitle{{More than telling a story: Transforming data into
  visually shared stories}}.
\newblock {\em IEEE computer graphics and applications\/} {35}, 5 (2015),
  84--90.
\newblock


\bibitem{moretti2013distant}
{Franco Moretti}. 2013.
\newblock {\em Distant Reading}.
\newblock Verso Books.
\newblock


\bibitem{north2000snap}
{Chris North} {and} {Ben Shneiderman}. 2000.
\newblock \showarticletitle{Snap-together visualization: a user interface for
  coordinating visualizations via relational schemata}. In {\em Proceedings of
  the working conference on Advanced visual interfaces}. ACM, 128--135.
\newblock


\bibitem{segel2010narrative}
{Edward Segel} {and} {Jeffrey Heer}. 2010.
\newblock \showarticletitle{{Narrative visualization: Telling stories with
  data}}.
\newblock {\em IEEE transactions on visualization and computer graphics\/}
  {16}, 6 (2010), 1139--1148.
\newblock


\end{thebibliography}

\end{document}